\documentclass[
  notitlepage,
  twocolumn,
  preprintnumbers,
  longbibliography,
  superscriptaddress,
  aps,
  prl,
  amsmath,
  amssymb,
]{revtex4-2}

\usepackage{physics}
\usepackage{tikz}
\usepackage{tkz-euclide}
\usetikzlibrary{positioning}
\usetikzlibrary{decorations.markings}
\usepackage{ifthen}
\usepackage{braket}

\usepackage{amsmath}
\usepackage{amsthm, amssymb}
\usepackage{slashed} 
\usepackage{graphicx}
\usepackage{xcolor}
\usepackage{bm}
\usepackage[
  pdfpagelabels,
  plainpages=false,
  bookmarks=true,
  colorlinks=true,
  linkcolor=red,
  urlcolor=blue,
  citecolor=blue
]{hyperref}
\usepackage{dsfont}
\usepackage{changepage}
\usepackage{array}

\def\a{\alpha}

\def\d{\delta}

\def\be{\begin{equation}}
\def\ee{\end{equation}}
\def\ba{\begin{align}}
\def\ea{\end{align}}
\def\>{\rangle}
\def\<{\langle}

\def\bea{\begin{eqnarray}}
\def\eea{\end{eqnarray}}
\def\>{\rangle}
\def\<{\langle}

\newcommand{\ZZ}{$\mathbb{Z}_2$}

\theoremstyle{definition}
\newtheorem{definition}{Definition}

\theoremstyle{remark}

\newtheorem{thm}{Theorem}
\tikzset{midarrow/.style={
    decoration={markings,
        mark= at position 0.625 with {\arrow{#1}} ,
    },
    postaction={decorate}
  }
}

\tikzset{beginarrow/.style={
    decoration={markings,
        mark= at position 0.375 with {\arrow{#1}} ,
    },
    postaction={decorate}
  }
}
\tikzset{endarrow/.style={
    decoration={markings,
        mark= at position 0.875 with {\arrow{#1}} ,
    },
    postaction={decorate}
  }
}
\begin{document}

\title{Fermionic Isometric Tensor Network States in Two Dimensions}

\author{Zhehao Dai}
\thanks{These two authors contributed equally.}
\affiliation{%
Department of Physics, University of California, Berkeley, CA 94720, USA
}%
\author{Yantao Wu}
\thanks{These two authors contributed equally.}
\affiliation{%
RIKEN iTHEMS, Wako, Saitama 351-0198, Japan
}
\affiliation{%
Department of Physics, University of California, Berkeley, CA 94720, USA
}%
\author{Taige Wang}
\affiliation{%
Department of Physics, University of California, Berkeley, CA 94720, USA
}%
\affiliation{Material Science Division, Lawrence Berkeley National Laboratory, Berkeley, CA 94720, USA}
\author{Michael P. Zaletel}
\affiliation{%
Department of Physics, University of California, Berkeley, CA 94720, USA
}%
\affiliation{Material Science Division, Lawrence Berkeley National Laboratory, Berkeley, CA 94720, USA}

\date{\today}

\begin{abstract}
We generalize isometric tensor network states to fermionic systems, paving the way for efficient adaptations of 1D tensor network algorithms to 2D fermionic systems. As the first application of this formalism, we developed and benchmarked a time-evolution block-decimation (TEBD) algorithm for real-time and imaginary-time evolution. The imaginary-time evolution produces ground-state energies 
for gapped systems, systems with a Dirac point, and systems with gapless edge modes to good accuracy. 
The real-time TEBD captures the scattering of two fermions and the chiral edge dynamics on the boundary of a Chern insulator.
\end{abstract}

\preprint{RIKEN-iTHEMS-Report-22}
\maketitle

Efficiently simulating quantum many-body systems is a central problem in 
computational physics.
Well-established numerical methods, including exact diagonalization (ED) and quantum Monte Carlo (MC), achieve great success in many problems, while still suffering from the accessible system size and the sign problem respectively.
For one-dimensional (1D) systems, density-matrix renormalization group (DMRG)~\cite{RevModPhys.77.259, white1992density} gives practically exact ground states for all gapped systems and good approximations for gapless systems.
The output of DMRG is a wavefunction written as a multiplication of matrices, hence the name matrix product states (MPS)~\cite{fannes1989exact,klumper1993matrix}.
Two-dimensional (2D) tensor network states (TNS) are generalizations of MPS; they (also known as PEPS) provide efficient representations for a large class of (if not all) ground states of local Hamiltonians~\cite{verstraete2004renormalization, cirac2021matrix}. 
However, the complexity of an \textit{exact} computation of any physical expectation value for a given 2D TNS scales exponentially with the system size. 
Impressive achievements on \textit{approximate} tensor contraction methods have been made with different balances of accuracy and cost \cite{simple_update, full_update_1, full_update_2, corner_transfer_matrix}, for both finite and infinite 2D TNS.
Another compromise between the 2D structure and the contraction efficiency are the tree tensor networks \cite{TTN}. 
Recently, Zaletel and Pollman sought a different path and proposed a restricted subclass of 2D TNS whose contraction is efficient by construction, the isometric TNS (isoTNS)~\cite{zaletel2020isometric}.
The key ingredient is the enforcement of the isometric condition in 2D, generalizing the canonical form of MPS. 
In addition to the reduced complexity due to the efficient contraction of the TNS overlap, this ansatz has the benefit of only needing to solve variational problems with condition number 1. 
This avoids the numerical instability in the optimization of the conventional TNS due to high condition numbers, which becomes more severe with higher bond dimension \cite{algorithm_finite_peps}.  
The improved stability opens up many possibilities, such as a DMRG algorithm in 2D~\cite{lin2021efficient} and real-time dynamics (local quench dynamics \footnote{For global quenches, the system quickly thermalizes and becomes a entanglement volume-law state, which cannot be described by a tensor network. Thus, we focus on local quenches here.}) to long time. 

In this work, we find that in a 2D TNS, the fermion sign structure is compatible with the isometric structure, which allows the notion of fermionic isoTNS (fisoTNS).
To accommodate the fermion statistics, the main challenge in 2D TNS is to account for any \textit{local} operator $c_i^\dagger c_j$ as a local TNS operation without a long Jordan-Wigner string \cite{JW}. 
People have introduced several equivalent~\cite{orus2014advances} formulations of 2D fermionic tensor network states (fTNS), including the virtual fermion ansatz~\cite{PhysRevA.81.052338,PhysRevA.80.042333,PhysRevB.81.245110}, the swap gate ansatz~\cite{shi2009graded,PhysRevB.81.165104}, and the Grassmann TNS \cite{gu2010grassmann}. 
We adopt the swap gate formalism in the main text, and provide a detailed derivation using virtual fermions in the Supplementary Materials (SM)~\cite{SM}.

We demonstrate a time-evolution block-decimation (TEBD) algorithm within the fisoTNS formalism.
As an example, we solve the ground states, via imaginary-time evolution, of various prototypical fermion systems.
For real-time dynamics, we see that fisoTNS has an advantage over traditional methods.  
We demonstrate two examples, including the dynamics of two fermions scattering in vacuum and low-energy edge excitations in a Chern insulator, up to a time scale comparable to the system size divided by the electron hopping rate, which we show to be very challenging for MPS algorithms \cite{WII,TDVP}. 
The system size we consider is also way beyond what ED can handle even after using the most advanced Krylov subspace techniques \cite{ED2,ED}. On the other hand, MC relies on analytic continuation to handle complex coefficients in real-time dynamics, which smear out any sharp feature in frequency \cite{AnalyticalContinuation,AnalyticalContinuation2,HubbardMC}.
The results we present are among the very few results \cite{fermion_dynamics_I} that simulate 2D fermion dynamics, whose success we attribute to improved stability due to the isometric form.

\begin{figure}[htb]
\begin{center}
\includegraphics[width=0.45\textwidth]{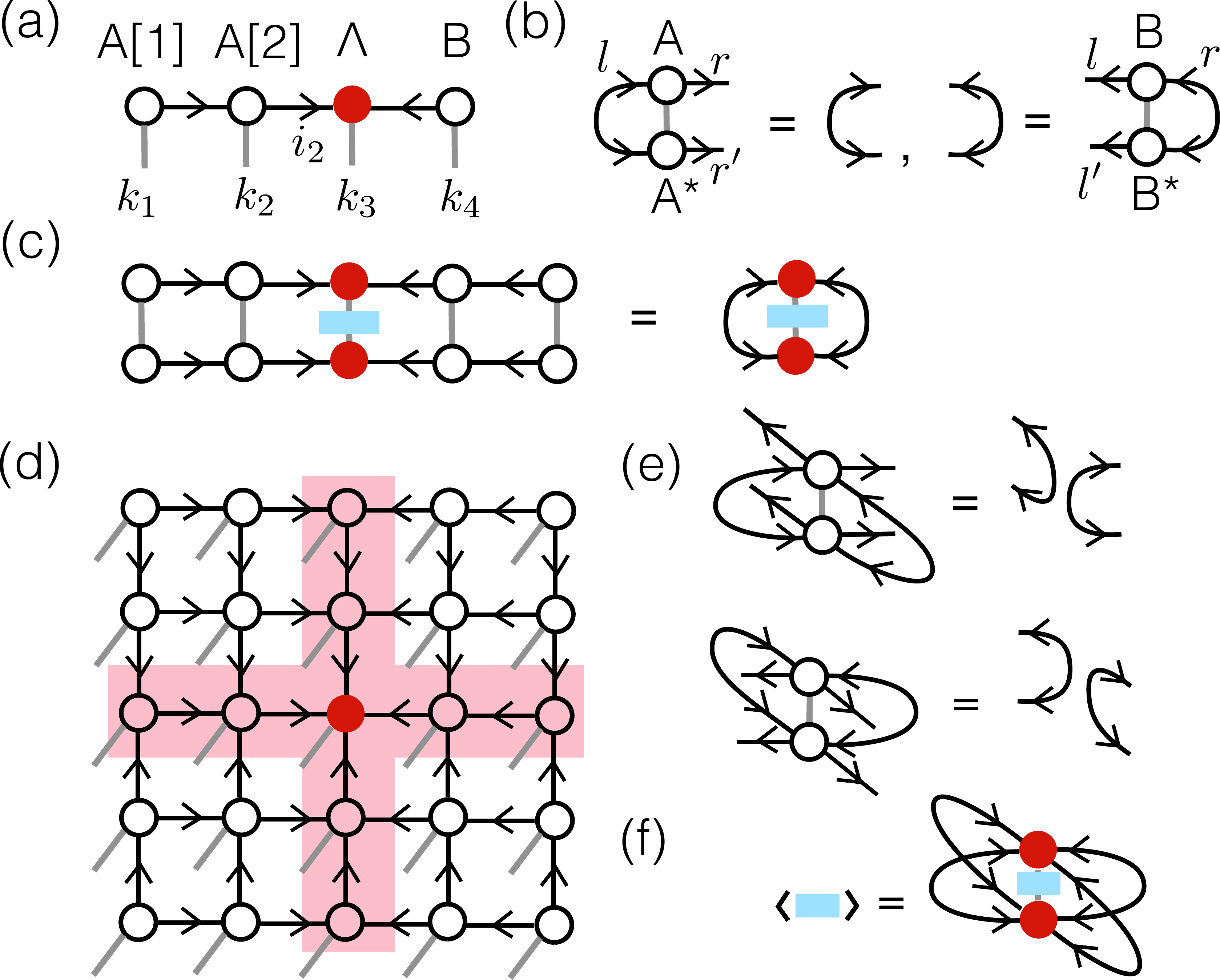}
\caption{Isometric tensor network states in 1D and 2D. (a) A matrix product state in isometric form. The red dot represents the orthogonality center. (b) Isometric condition for tensors on the left (right) of the center $A$ ($B$). A line without a circle on it represents identity. (c) The expectation of any physical operator (blue square) on the orthogonality center equals the contraction of the operator with the central tensor and its conjugate. (d) 2D isometric tensor network. The red dot represents the orthogonality center. On the shaded region, the orthogonality hypersurface, all arrows point in. (e) Isometric condition for tensors in the lower-left and the upper-right quadrant. (f) The same as (c) but in 2D.}
\label{FigIso}
\end{center}
\end{figure}

\textit{Isometric condition.} 
We review the isometric condition for bosonic TNS. 
Consider an isometric MPS: $\psi^{k_1,\dots,k_N} = \sum_{\{i_n\}} A[1]^{k_1}_{i_1} A[2]^{k_2}_{i_1,i_2} \cdots \Lambda[n_c]_{lr}^{k_{n_c}}\cdots B[N]^{k_N}_{i_{N-1}}$ for some site $n_c$ (Fig.~\ref{FigIso}a).
The isometric condition for tensors left (right) to $n_c$ is $\sum_{l,k}A^{k*}_{lr'}A^{k}_{lr} = \delta_{rr'}$ ($\sum_{r,k}B^{k*}_{l'r}B^{k}_{lr} = \delta_{ll'}$)   (Fig.~\ref{FigIso}b). 
Under this condition, the expectation value of any physical operator $O_{k,k'}$ at $n_c$ takes a simple form $\sum_{l,r,k,k'}\Lambda^{k*}_{lr}O_{k,k'}\Lambda^{k'}_{lr}$ (Fig.~\ref{FigIso}c). 
In 1D, any MPS can be put into the isometric form without increasing the bond dimension~\cite{RevModPhys.77.259}. 

2D TNS wavefunction is constructed in a similar manner (Fig.~\ref{FigIso}d).
The generalized \textit{isometric} constraint is the following~\cite{zaletel2020isometric}.
First, we assign arrows on each tensor leg.  
The arrows on the physical legs always point into the tensors. 
On the virtual legs, all horizontal (vertical) arrows point toward a chosen column (row). 
The chosen column and row together form the {\it orthogonality hypersurface} (OH) (Fig.~\ref{FigIso}d red shaded), crossing at the {\it orthogonality center} (red dot in Fig.~\ref{FigIso}d). 
Then we require each tensor, viewed as a map from the outgoing legs to the incoming legs, to be an isometry. 
For example, for tensors in the lower-left quadrant:
$\sum_{l,d,k}A^{k*}_{lrdu}A^{k}_{lr'du'} = \delta_{rr'}\delta_{uu'}$ (Fig.~\ref{FigIso}e).
With this isometric condition, computing correlation functions on the OH reduces to 1D tensor contractions ~\cite{zaletel2020isometric}.
In particular, expectation values of physical operators on the orthogonality center only depend on a single tensor (Fig.~\ref{FigIso}f).
Unlike in 1D, 2D isoTNS is a strict subclass of general TNS; nonetheless, it still includes all bosonic topological orders with gappable boundaries~\cite{stringnet_isotns}.

\begin{figure}[htb]
\begin{center}
\includegraphics[width=0.45\textwidth]{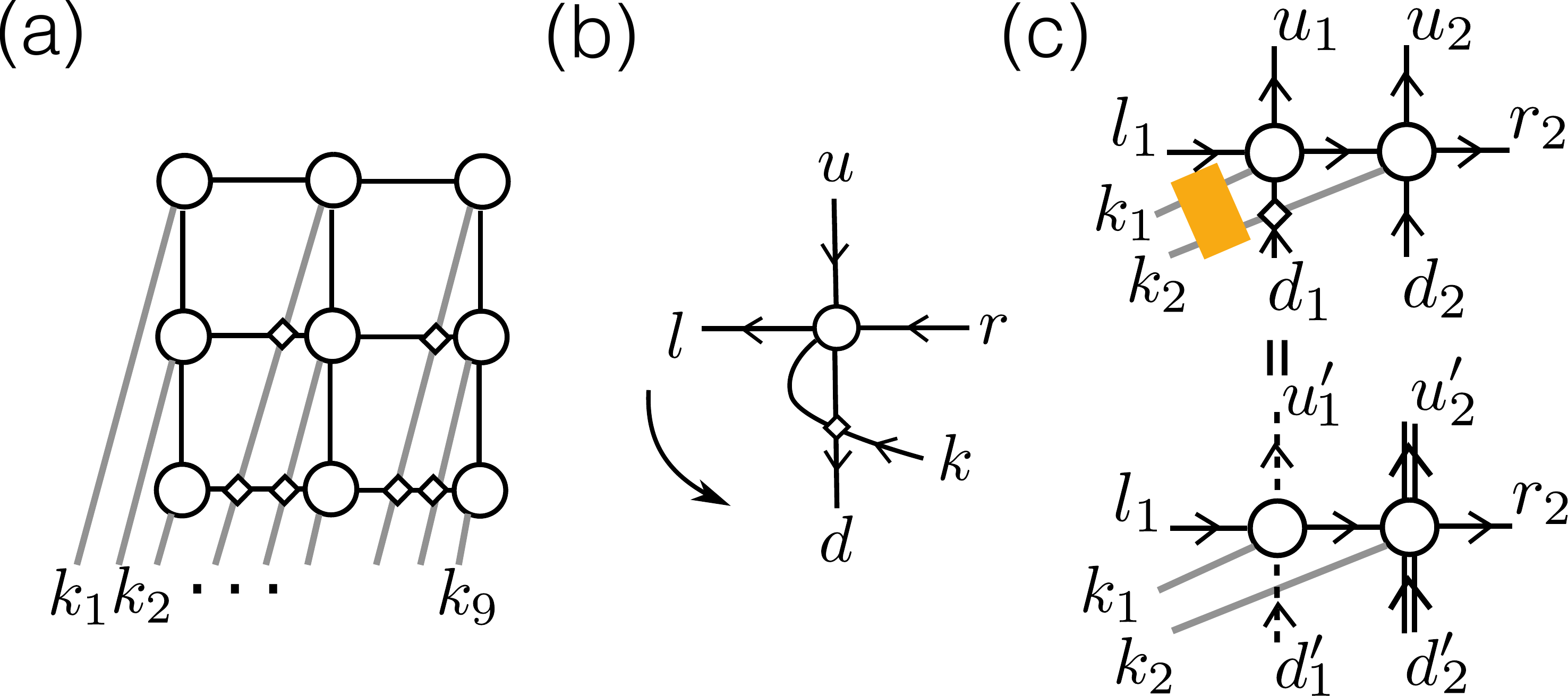}
\caption{(a) Definition of the fermionic wavefunction in the swap gate convention. An extra sign (diamond) is inserted at every crossing. (b) In the upper-right quadrant of a fisoTNS, we require $(-1)^{P_kP_d} A$ to be an isometric tensor. The minus sign comes from putting the outgoing legs next to each other. (c) After the action of a brick wall circuit, we can write the resulting state as a new fisoTNS by splitting the product of two tensors, the unitary and the swap gate back into two isometric tensors. Here we choose a particular way of splitting the tensor where the original vertical legs are assigned to the second tensor and the left tensor has trivial up and down legs (dashed lines)}
\label{Figswapiso}
\end{center}
\end{figure}

\textit{Fermionic isometric tensor network states}-- To account for the fermion sign structure, it is necessary to keep track of the fermion parity of each site explicitly. We require every tensor to be $\mathbb{Z}_2$ symmetric, and, therefore, both physical and virtual legs to be labeled with a $\mathbb{Z}_2$ parity charge \cite{hauschild2018efficient}. In the swap gate convention, the physical legs are drawn to extend outside the system and the endpoints are ordered as $k_1, k_2,\dots k_n$ counterclockwise, as in Fig. \ref{Figswapiso}(a). Then the fermion wavefunction in the basis of $c_1^{\dagger k_1}\dots c_n^{\dagger k_n}|\mathrm{vac}\>$ is defined as the contraction of tensors on each site and additional `swap gates' at each crossing of physical legs and virtual legs (Fig.~\ref{Figswapiso}(a)).
The swap gates are simply a sign $(-1)^{P_i P_j}$, where $P_i, P_j = 0, 1$ are parity of the two crossing legs, e.g. $(-1)^{P_k P_l}$ when a physical leg crosses a left leg~\cite{shi2009graded,PhysRevB.81.165104}. The virtual indices can take any integer value up to any chosen bond dimension. (See SM~\cite{SM} for a self-contained review.)

Similar to the bosonic case, we define fisoTNS by requiring that the double-layer tensor network representing $\<\psi|\psi\>$ trivially contracts to the center.
\begin{definition} (fisoTNS)
  Given a 2D fTNS with isometric arrows on its tensor legs, it is a {\it fermionic isoTNS} if I) $(-1)^{P_kP_d} A$ is isometric for any tensor $A$ to the upper-right of the orthogonality center; and II) all other tensors are isometric as in the bosonic case.  
\end{definition}
The isometric rule for a tensor is the same as the bosonic case when the \textit{outgoing} legs of the tensor are next to each other. 
This condition is true in every quadrant except the upper-right quadrant, where we need to exchange the order of the physical and the down leg to make it happen, hence the extra sign.
With this definition, we can state a key theorem of fisoTNS~\cite{SM}.
\begin{thm}
\label{thm:overlap}
For any two fisoTNS $|\psi\>$ and $|\psi'\>$ that have the same tensors 
everywhere except at the orthogonality center, with $\hat{\Lambda}_{(x_0,y_0)}$ and $\hat{\Lambda}'_{(x_0,y_0)}$ for each respectively,  $\<\psi|\psi'\> = \sum_{l,r,d,u,k}\Lambda^{*k}_{l,r,d,u}\Lambda'^{k}_{l,r,d,u}$.
\end{thm}
Here all the swap gates in $\bra{\psi}$ and $\ket{\psi'}$ delicately cancel, which guarantees the contraction of fisoTNS to be efficient.

\textit{Representability.} As shown in Fig.~\ref{Figswapiso}c, a local two-site unitary gate affects only two tensors. If a brick wall circuit acts on an fisoTNS, we can always find an fisoTNS representation of the resulting state by first grouping the tensors connected by gates, absorbing the \textit{unitary} gates, and splitting the tensor back to isometric tensors on each site
(See SM for details.). 
This argument shows that if a state admits an fisoTNS representation, any other state in the same phase also has an fisoTNS representation.
Thus, all topologically trivial insulators, interacting or not, have an accurate fisoTNS representation.

\begin{figure}[htb]
\begin{center}
\includegraphics[width=0.45\textwidth]{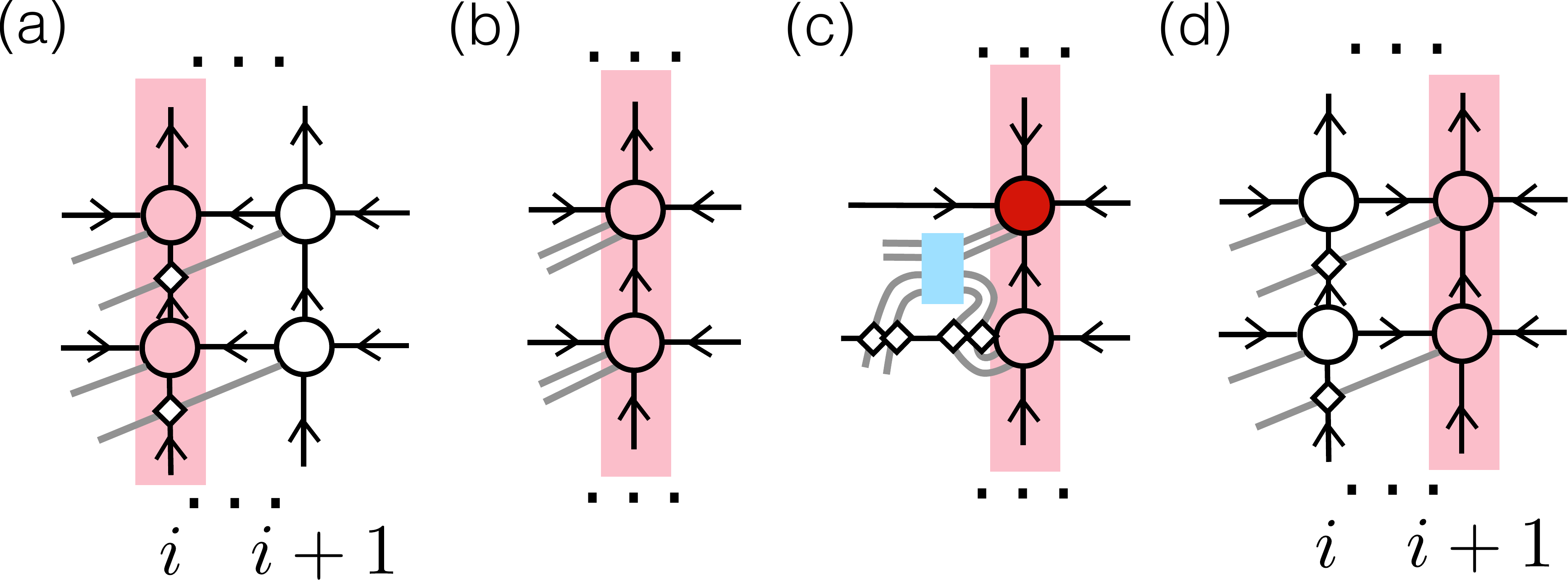}
\caption{TEBD in fisoTNS.
(a) Column $i$ is the OC.
(b) Contraction of column $i$ and $i+1$ into one column. 
(c) Apply the TEBD gates on the OC using the MPS TEBD method.
(d) After the MM algorithm, column $i+1$ is now the OC.
}
\label{fig:TEBD}
\end{center}
\end{figure}

\textit{Algorithms.}
As a first application, we develop a TEBD algorithm \cite{TEBD, iTEBD} for 2D lattice fermion systems (Fig. \ref{fig:TEBD}).  
We will only speak of the orthogonality column (OC), with the orthogonality row always on the top. 
For an fisoTNS with OC column $i$, we first contract column $i$ and $i+1$ into one combined column with two physical sites on each tensor and apply $O_{i,i+1}$ to it using MPS methods. 
Then we split the combined column back into column $i$ and $i+1$, but with column $i + 1$ now as the OC. 
The splitting aims to variationally maximize the overlap before and after the split while preserving the isometric constraints.  
This is solvable with Riemannian optimization techniques \cite{Riemannian_optimization_isoTN, algorithm_MERA}. 
Here we initialize the variational problem with the Mose's Move (MM) algorithm, introduced in \cite{zaletel2020isometric} and generalized to symmetric tensors in \cite{SM}. 
The MM and the Riemannian optimization is the computational bottleneck of the algorithm, whose computational complexity is $O(\chi^7f^3)$, where the bond dimension is $f\chi$ on the OH and $\chi$ elsewhere.    
Repeating these steps, one can perform real or imaginary time evolution of a 2D fermion Hamiltonian following the Trotter-Suzuki decomposition~\cite{trotter}.  
The fermion signs necessary for TEBD are {\it all} shown in Fig. \ref{fig:TEBD} as swap gates.  

\textit{Ground state results.} To benchmark the TEBD algorithm, we study four prototypical fermionic systems on a square lattice: a sublattice band insulator (Insulator), a Dirac semimetal (Dirac), 
a $\phi = 2\pi/3$ Hofstadter model with total Chern number = 1 (Chern), and a mean-field $p+ip$ superconductor ($p+ip$ SC),
\begin{equation}
    H = \sum_{\langle ij \rangle} t_{ij} c_i^{\dagger} c_j + \eta_{ij} c_i^{\dagger} c_j^{\dagger} + \mathrm{h.c.} + U \hat{n}_i \hat{n}_j - \sum_{i} \mu_i c_i^{\dagger} c_i\\
\end{equation}
In all systems, $|t_{ij}| = 1$, and in $p+ip$ SC, $|\eta_{ij}| = 2$. 
The phases of $t_{ij}$ ($\eta_{ij}$) set the flux (pairing symmetry) of the Hofstadter model ($p+ip$ SC). 
All systems are non-interacting with $U = 0$, except for the Chern band where $U=0.5$ is also explored. 
We set $\mu_i = \pm 1$ for the two sublattices for the insulator, and $\mu_i = -2$ in the $p+ip$ SC. 
For the Chern band (Dirac), we set a uniform $\mu_i$ such that the system is $2/3$ filled (half-filled).


\begin{figure}[htbp]
\begin{center}
\includegraphics[width=0.43 \textwidth]{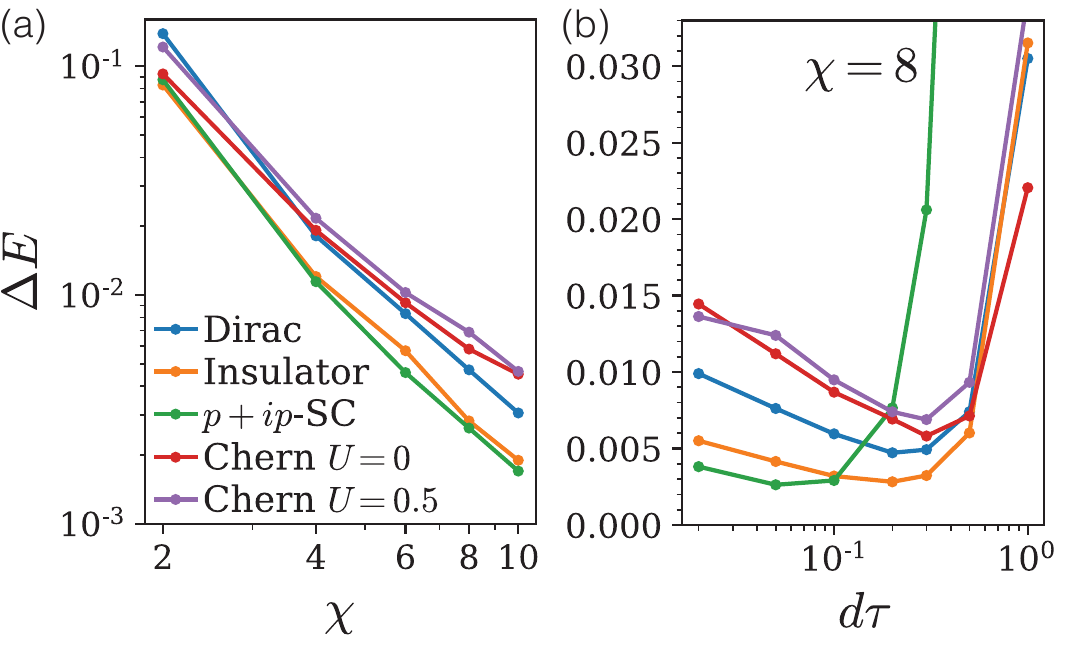}
\caption{Error in ground state energy density. (a) is presented with the optimal $d\tau$. The system size is $9 \times 9$ for Chern and $10 \times 10$ for Dirac, Insulator, and $p+ip$ SC. 
$f = \chi$. 
The MM error is typically $10^{-4}$ or $10^{-5}$, comparable with the truncation error of the MPS TEBD within the OC.  
}
\label{fig:energy}
\end{center}
\end{figure}



\begin{figure*}[htbp]
\begin{center}
\includegraphics[width=0.9\textwidth]{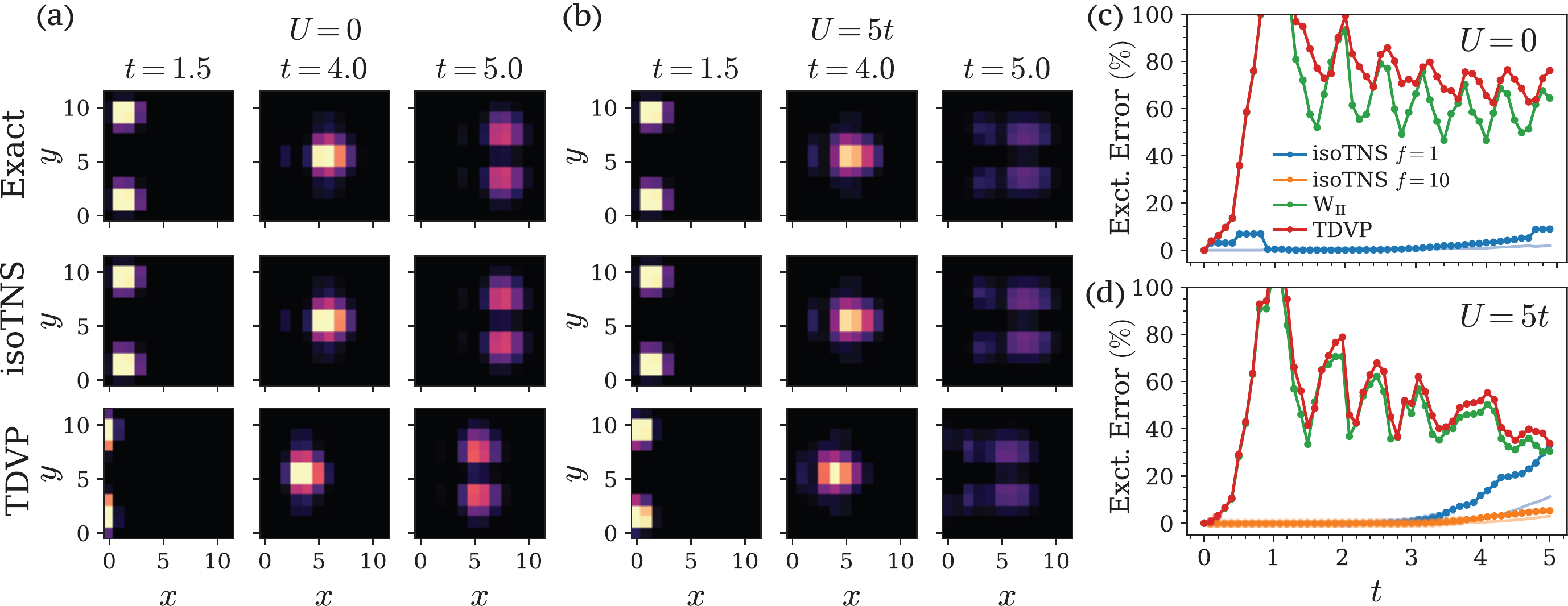}
\caption{Real-time evolution of two fermions scattering with nearest neighbor hopping on a $12 \times 12$ square lattice without (a) and with (b) interaction. The two fermions are initialized at the top-left and bottom-left corner. We show the ED, fisoTNS, and 1D TDVP results from top to bottom. From left to right, we show three time slices, before, during and after the collision. 
We show the error of various methods in (c-d), including fTNS with different configurations and 1D W$_{\mathrm{II}}$ and TDVP methods. The solid lines show the excitation error and the dashed lines show the background error (see main text for definition).}
\label{fig:scatter}
\end{center}
\end{figure*}

\begin{figure}[htbp]
\begin{center}
\includegraphics[scale=0.5]{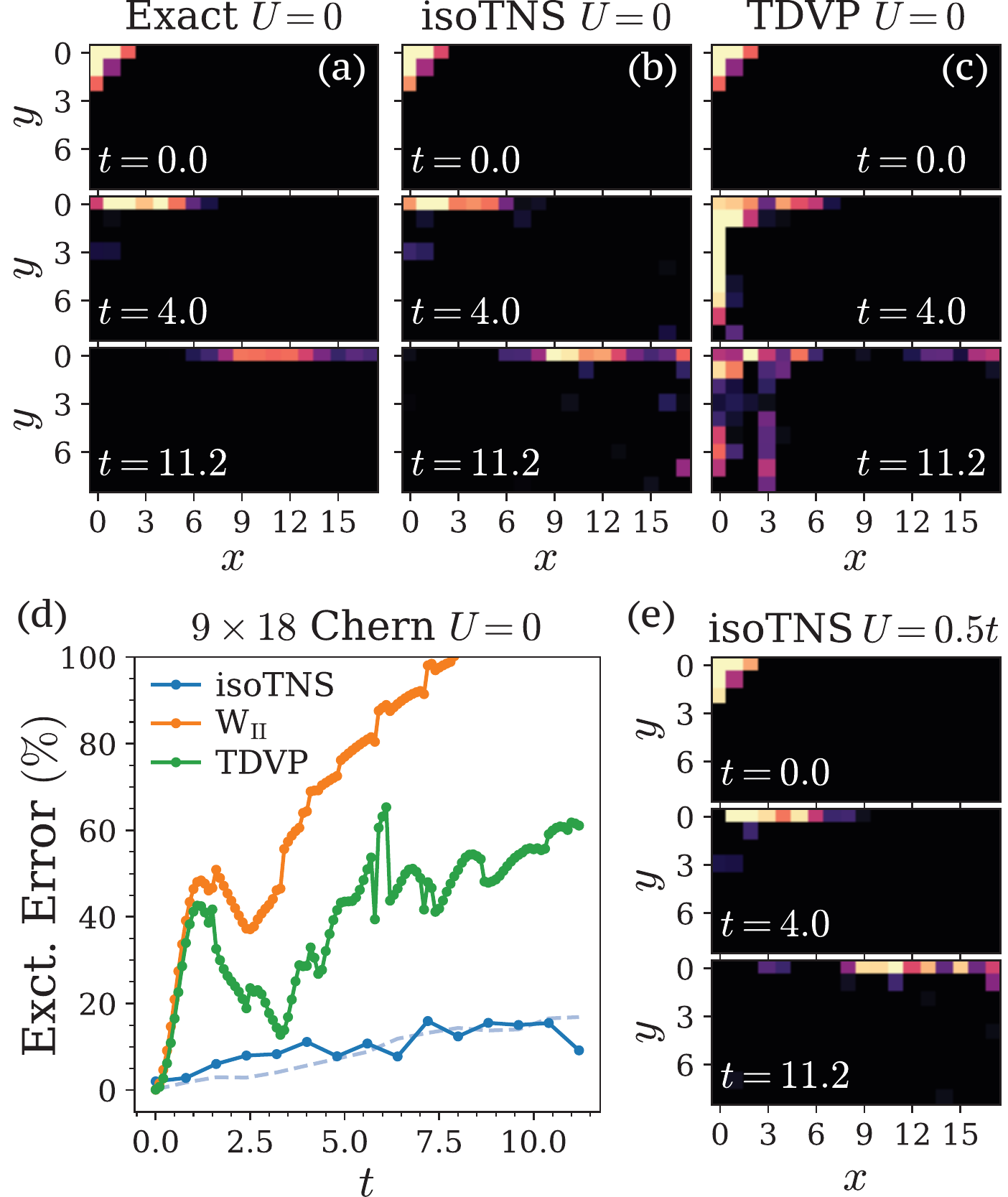}
\caption{Real-time evolution of the $\phi = 2\pi/3$ Hofstadter model starting with a single-site excitation at the top-left corner of a $9 \times 18$ system. (a-c, e) Number density relative to the ground state. (d) Errors of the excitation sites when $U=0$.}
\label{fig:time}
\end{center}
\end{figure}

We perform 2nd-order imaginary-time TEBD to find the ground state energies, compared with the exact values~\footnote{ 
To benchmark the Chern band with finite interaction, we use $\mathsf{U}(1)$-conserved density matrix renormalization group (DMRG) with snake-winding and bond dimension $\chi = 2400$.}. 
In Fig.~\ref{fig:energy}, we show the error in ground state energy density at various bond dimensions $\chi$ and Trotter time $d\tau$. 
This error consists of accumulated MM error and second order trotter error: $\Delta E = a\ \epsilon_{\mathrm{MM}}/d\tau + b\ d\tau^{4}$ \cite{zaletel2020isometric}, where the MM error $\epsilon_{\mathrm{MM}}$ decreases as $\chi$ increases. For a fixed $\chi$, there is a system-dependent optimal $d \tau$, which gives the lowest ground-state energy (Fig.~\ref{fig:energy} b)~\footnote{The optimal $d\tau$ goes to zero as the bond dimension increases.}.
These results provide information on the representibility of tensor networks with isometric constraints, in particular the ability of sequential quantum circuits to represent fermionic ground states \cite{sequential_circuits}. 
We also note that these results can be improved with a 2D fermion DMRG algorithm \cite{lin2021efficient} free of Trotter error in the future. 

\textit{Real-time dynamics.} fisoTNS reveals its true advantage for real-time dynamics. 
We start by benchmarking a two-fermion scattering problem with nearest neighbor hopping $t_{ij} = 1$ and interaction $U$. 
We restrict ourselves to two particles so that the result can be benchmarked with exact diagonalization (ED). 
As shown in Fig.~\ref{fig:scatter}, fisoTNS reproduces the number density on each site during the scattering. 
MPS methods, including the time-dependent variational principle (TDVP) \cite{TDVP} and the W$_{\mathrm{II}}$ method \cite{WII}, fail to capture even the beginning of the dynamics. 
The main reason is that the snake-winding nature of the MPS prevents the number density to move across the winding~\footnote{Future improvements on TDVP and W$_{\mathrm{II}}$ that specialize on 2D problems, including subspace expansion, might help to improve their performance. Nevertheless, we are not aware of any published works on these improvements.}. 
As for PEPS, fermion dynamics is considered a very challenging task, as evidenced by the few number of published studies in the literature, which evolve the dynamics to relatively short times \cite{fermion_dynamics_I}\footnote{Our comment here is of course not to discredit any of the PEPS fermion dynamics simulations in the literature.
The previous results are very impressive on their own. 
We just would like to note the general challenge of this problem.}. 

To quantify the error in the real-time dynamics, we define the excitation error to be the total error on the sites that support the first $60\%$ of the added particle number, and the background error to be the error per site on the rest of the sites \footnote{To be precise, At each time step $t$, we rank sites by their particle number density and select the top $n$ sites whose total particle count reaches $1.2$, representing $60\%$ of a two-particle excitation. These are termed ``excited sites.'' The normalized excitation error is then calculated as the sum of the absolute deviations in particle densities within these excited sites from their true values, divided by $1.2$, which serves as the hypothetical maximum error. For non-excited sites, we compute the average absolute deviation to gauge the background error. This background error is normalized by multiplying the number of excited sites and then dividing by $1.2$, allowing for a direct comparison with the normalized excitation error.}. 
This way, for local excitations, the error does not scale with the system size despite random background noise. 
With $\chi=12, f=10$, the error is within $5 \%$ towards the end of the simulation.

We now consider a truly challenging problem, the chiral propagation of a fermion on the edge of a Chern insulator, which has not yet been demonstrated in an interacting system \footnote{A bosonic version has been demonstrated in Ref.~\onlinecite{bosonFCI}}. 
We start with a ground state $|\mathrm{GS}\>$ of a Hofstader model described in the previous section, add an electron at the corner, $\ket{\mathrm{ext}} = c^{\dagger} \ket{\mathrm{GS}}$, and then evolve the excited state for time $t$, $\ket{t} = e^{-iHt} \ket{\mathrm{ext}}$. 
We show the number density of each site relative to the ground state $n_i = \expval{\hat{n}_i}{t} - \expval{\hat{n}_i}{\mathrm{GS}}$ in Fig.~\ref{fig:time}. 
For both the free and interacting cases, the excitation propagates along the edge without leaking into the bulk even on the time scale of system size divided by hopping rate. 
MPS based methods fail to capture the beginning of dynamics.
Comparing Fig.~\ref{fig:time}(a) and (b), fisoTNS reproduces not only the velocity of the edge mode but also subtle features of the density profile (e.g. at $t = 4$). 
The error is shown in Fig.~\ref{fig:time}(d), within $10\%$ toward the end of the simulation.
This simulation demonstrates the ability of 2D TNS to qualitatively capture the edge properties of a finite chiral system, in complementary to the recent studies of the TNS's ability to \textit{approximately} capture the bulk properties of infinite chiral systems \cite{chiral_iPEPS_I, chiral_iPEPS_II}. 

\textit{Open problems.}
The fisoTNS formalism opens many interesting questions. 
First, it would be interesting to explore its representability of fermionic topological order \cite{fermion_TO,fermion_TC,fTNS_TP}. 
Given that bosonic isoTNS can exactly represent all string-net fixed-point states \cite{stringnet_isotns}, it would be very interesting to show that fisoTNS similarly can represent all gappable fermionic topological states in 2D. 
Along the same line, it would be interesting to find the entanglement criterion \footnote{For example, for MPS, the entanglement criterion is the Renyi entanglement entropy being area-law.} for a state representable by fisoTNS.
Because of the 2D nature and the fact that the OH partition the systems into four disjoint parts, the entanglement criterion will likely be a multipartite one \cite{h_Karthik}. 
In fact, from a numerical perspective, the fisoTNS is also a good tool to compute the 4-party properties of a fermion system, as all information of the 4-party entanglement is contained in the tensors on the OH, a much easier object for computation than the full 2D wavefunction. 
Another direction around the representability question is to develop a Gaussian formalism \cite{PhysRevA.81.052338, GfTNS_chiral, TN_Fermi_Surface} for fisoTNS so that understanding can be gained for the relatively simple free fermion systems. 
From the algorithmic perspective, for example, it should be possible to develop a genuinely 2D DMRG algorithm for fisoTNS, which would remove the $d\tau$ dependence in the imaginary-time algorithm. 
Another possibility lies in simulating finite temperature many-body fermion systems by generalizing the METTS \cite{METTS, algorithm_METTS} algorithm to 2D. 
This is especially suited for isoTNS, because, unlike a generic 2D TNS, sampling of an isoTNS is efficient \cite{perfect_sampling}.

\textit{Conclusion.}
We generalized isoTNS to fermionic systems, which allows for efficient adaptations of 1D MPS algorithms to 2D fermionic systems.
The 2D TEBD algorithm we developed as the first application produces ground states and real-time dynamics of low-lying states with good accuracy. 
The potential in simulating local-quench dynamics for a long time is particularly interesting. 
This task is generally believed to be very challenging. 
Our study shows that fisoTNS has a clear advantage over existing methods.
With fisoTNS, it is straightforward to measure any time-dependent correlation functions of local operators, directly compare our results with experiments like ARPES, STM, neutron scattering, and transport, and even measure excitations that experiments can hardly detect.

\textit{Acknowledgement.} 
We thank Sheng-Hsuan Lin, Johannes Hauschild, and Peter Lunts for helpful discussions. YW acknowledges financial support from the RIKEN iTHEMS fellowship. This work was partly supported by the Gordon and Betty Moore foundation. 
MZ was supported by the U.S. Department of Energy, Office of Science, Basic Energy Sciences, under Early Career Award No. DE-SC0022716.
TW is supported by the U.S. Department of Energy, Office of Science, Office of Basic Energy Sciences, Materials Sciences and Engineering Division under Contract No. DE-AC02-05-CH11231 (Theory of Materials program KC2301 and van der Waals Heterostructures Program KCWF16). TW is also supported by the Heising-Simons Foundation, the Simons Foundation, and NSF grant No. PHY-2309135 to the Kavli Institute for Theoretical Physics (KITP).
The code is implemented based on the TenPy TNS code base \cite{hauschild2018efficient}. This research used the Lawrencium computational cluster resource provided by the IT Division at the Lawrence Berkeley National Laboratory (Supported by the Director, Office of Science, Office of Basic Energy Sciences, of the U.S. Department of Energy under Contract No. DE-AC02-05CH11231).

\bibliography{ref.bib}

\clearpage
\onecolumngrid
\appendix

\section{Virtual fermion ansatz}
\label{appendix:virtualfermion}

\begin{figure}[htb]
\begin{center}
\includegraphics[width=0.45\textwidth]{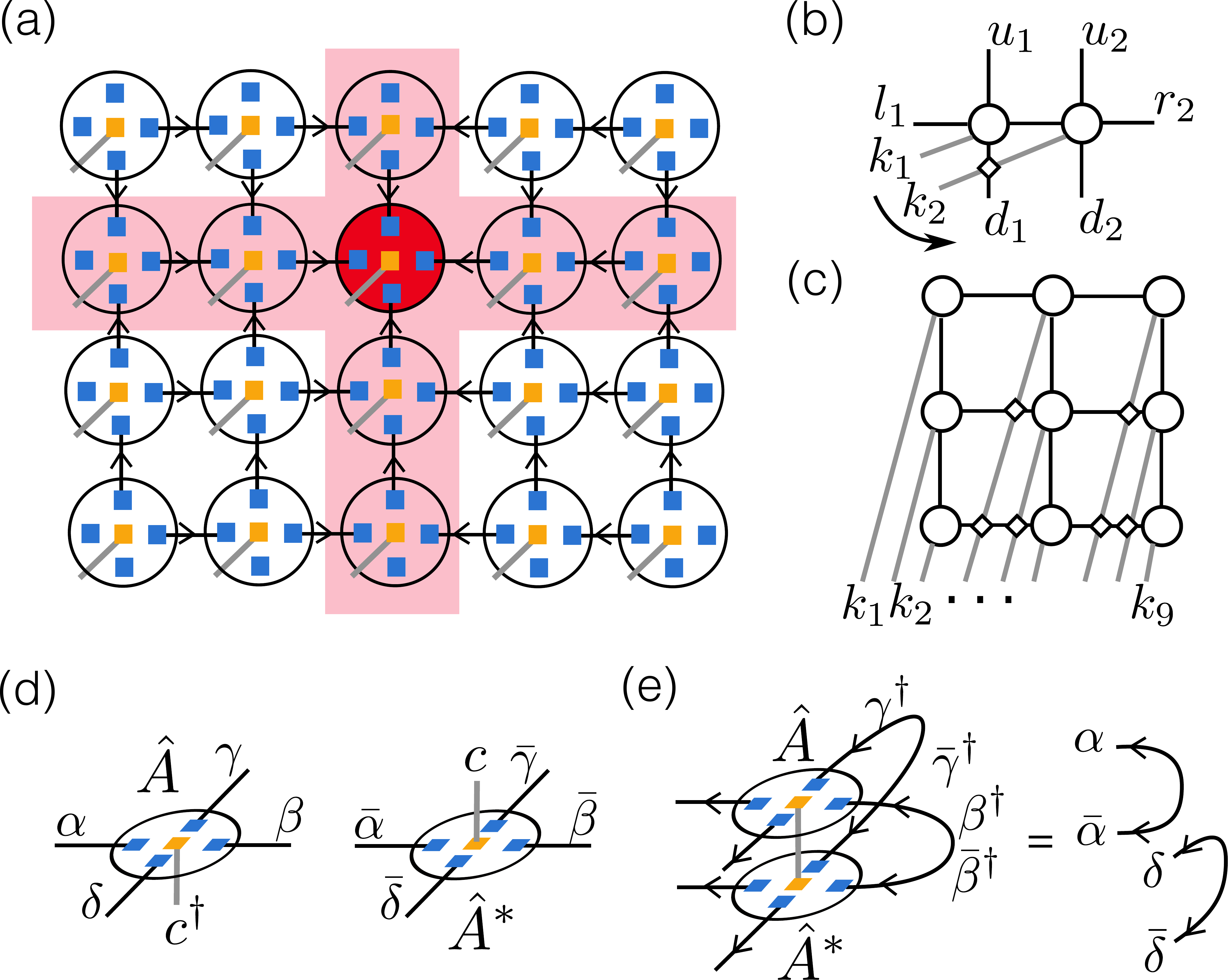}
\caption{(a) Fermionic isometric tensor network states in the virtual-fermion formalism. The arrows point to the orthogonality center as in the case of bosonic isoTNS.  The 4 blue squares on each site represent 4 virtual complex fermions. Orange squares represent physical fermions. Each tensor is promoted into an operator by attaching the physical creation operator and the virtual fermion annihilation operator. The tensors act on the virtual-fermion state of one bell pair per bond to create the physical state. (b) To contract two fermionic tensors and order the fermion operators counterclockwise starting from $k_1$, we need to do the tensor contraction with an additional swap gate $(-1)^{k_2D_1}$ (the small square). (c) In the swap-gate convention, the wavefunction is given by the contraction of all tensors and swap gates at the crossings. (d) Fermion-attachment of the tensor $A$ and $A^*$. (e) Isometric condition for fisoTNS in the upper-right quadrant.}
\label{FigfIso}
\end{center}
\end{figure}

We now introduce an fTNS ansatz that explicitly attaches auxiliary fermions to virtual bonds (a variation of the ansatz in Ref.\cite{PhysRevA.81.052338}).
\begin{align}
    |\psi\> &= \< \prod_{(x,y)} \hat{A}_{(x,y)}\prod_{(x,y)}\hat{V}_{(x,y)}\>_\text{aux}|vac\rangle,\label{EqfPEPS}\\
    \hat{A}_{(x,y)} &= A^k_{lrdu} c_{(x,y)}^{\dagger k}\delta_{(x,y)}^{D}\beta_{(x,y)}^{R}\gamma_{(x,y)}^{U}\alpha_{(x,y)}^{L},\label{Eqfermiontensor}\\
    \hat{V}_{(x,y)} &= (1+\beta_{(x,y)}^{\dagger}\alpha_{(x+1,y)}^{\dagger})(1+\delta_{(x,y+1)}^{\dagger}\gamma_{(x,y)}^{\dagger}),
\end{align}
where we assume a physical Hilbert space of one complex fermion per site, $c_{(x,y)}$, and we omit indices of $\hat{A}_{(x,y)}$ for simplicity. 
Each tensor $A^{k}_{lrdu}$, $k\in\{0,1\}$, $l,r,d,u\in\{0,\dots,\chi-1\}$, is required to be $\mathbb{Z}_2$-graded, where the $\mathbb{Z}_2$ symmetry corresponds to the fermion parity conservation. 
Namely, for each virtual index $l,r,d,u$, we assign a parity label $P_l, P_r, P_d, P_u = 0, 1$. For simplicity, we write these parity labels simply as capital letters $L, R, D, U$ correspondingly. And We require $A^{k}_{lrdu} = 0$ whenever $k+L+R+D+U$ is odd.
On each site, we assign 4 auxiliary complex fermions $\alpha,\beta,\gamma, \delta$, on the left, right, up, and down virtual bonds respectively (Fig.~\ref{FigfIso}(a)).
The tensors are promoted to fermionic operators according to the rule that whenever the corresponding index has a parity label $1$, an auxiliary (physical) fermion annihilation (creation) operator is attached (Eq.~\ref{Eqfermiontensor}).
Auxiliary fermion operators on each site are ordered \textit{counterclockwise}.
As for usual TNS, whenever two sites are connected, the corresponding two virtual indices are contracted.
Furthermore, the two auxiliary fermion annihilation operators attached to the bond are contracted with the bond operator $(1+\beta_{(x,y)}^{\dagger}\alpha_{(x+1,y)}^{\dagger})$ for the horizontal and $(1+\delta_{(x,y+1)}^{\dagger}\gamma_{(x,y)}^{\dagger})$ for the vertical bond.
The action of the bond-operators and the site-operators on the vacuum of the physical fermions, followed by the contraction of all virtual fermion operators in their vacuum gives the desired physical state.
For example, the fTNS with bond dimension one, where $\hat{A}_{(x,y)} = c^{\dagger}_{(x,y)}\beta_{(x,y)}$ for even $x$ and $\hat{A}_{(x,y)} =  c^{\dagger}_{(x,y)}\alpha_{(x,y)}$ for odd $x$, gives the fully filled state of one fermion per site. 
We may allow more than one pair of entangled fermions for each virtual bond in the definition of fTNS, but the resulting ansatz can always be rewritten in the form of Eq.~\ref{EqfPEPS}~\cite{SM}.

This ansatz avoids the problem of the nonlocal string discussed before.
Since each site-operator $\hat{A}_{(x,y)}$ is parity-even, the fermion annihilation (creation) operator $c_{(x,y)}$ ($c^{\dagger}_{(x,y)}$) \textit{commutes} with all site-operators on sites other than $(x,y)$. Thus the action of nearest-neighbor hopping affects only the two tensors on the corresponding sites. 

The ansatz in Eq.~\ref{EqfPEPS} may seem complicated, but it does not increase the complexity of any tensor network algorithm. 
The $\mathbb{Z}_2$ symmetric tensors are routinely used for $\mathbb{Z}_2$ symmetric boson models; the only thing the auxiliary fermions do is to give extra minus signs when contracting tensors.
For example, if we want to replace the two tensors in Fig.~\ref{FigfIso}(b) by a single tensor $\hat{\Theta} = \Theta^{k_1k_2}_{l_1r_2d_1d_2u_1u_2}c_1^{\dagger k_1}c_2^{\dagger k_2}\delta_1^{D_1}\delta_2^{D_2}\beta_2^{R_2}\gamma_2^{U_2}\gamma_1^{U_1}\alpha_1^{L_1}$ without changing the many-body state, we need
$\hat{\Theta} = \sum_{r1=l2}\< \hat{A}[1] \hat{A}[2] (1+\beta^{\dagger}_1\alpha^{\dagger}_2)\>_{\beta_1,\alpha_2}$,
where we contract the auxiliary fermion $\beta_1$ and $\alpha_2$ in their vacuum.
By direct calculation, (indices omitted unless necessary)
\begin{equation}
\label{eq:sign}
\Theta = A[1]A[2](-1)^{k_2D_1}.
\end{equation}

In fact, extra factors like $(-1)^{k_2D_1}$ can be deduced pictorially without auxiliary fermions (Fig.~\ref{FigfIso}(b)):
The rule is whenever two fermion legs with parity labels $I$ and $J$ cross each other, an extra \textit{swap gate}, $(-1)^{IJ}$ is applied~\cite{PhysRevB.81.165104}.
Using swap gates, the fermion wavefunction can be written in the Fock space basis as the regular contraction of the tensors $A_{(x,y)}$ with additional swap gates applied at the crossings as we extend the physical legs outside (Fig.~\ref{FigfIso}(c)).
In extending the physical legs, we can freely deform them~\cite{PhysRevB.81.165104} 
without changing the wavefunction as long as the order of endpoints is fixed. The latter corresponds to the order of fermions of the Fock space basis.
The general equivalence between the swap-gate method and auxiliary-fermion ansatz is discussed in~\cite{orus2014advances}.
A self-contained proof for our proposal is given in supplemental materials (SM)~\cite{SM}.
In practice, we only need to keep track of fermion signs that arise from local tensor operations instead of actually computing the wavefunction. 

\textit{Isometric condition for fermions} 
As in bosonic isoTNS, the isometric condition for fermions should make trivial the contraction between a state and its conjugate.
One may guess that this is just the isometric condition at the tensor level.
This is almost correct. In fact, the isometric condition we define below is equivalent to that $A$ is an isometry in the upper-left, lower-left, and lower-right quadrants, but $(-1)^{kD}A$ be an isometry in the upper-right quadrant.
The asymmetry among the 4 quadrants is related to how we order fermion operators in Eq.~\ref{Eqfermiontensor}.

More precisely, we {\it define} a set of conjugate tensor operators
$\hat{A}_{(x,y)}^* = A^{*k}_{l'r'd'u'} \bar{\alpha}_{(x,y)}^{L'}\bar{\gamma}_{(x,y)}^{U'}\bar{\beta}_{(x,y)}^{R'}\bar{\delta}_{(x,y)}^{D'}c^{k}$,
where $\bar{\alpha}_{(x,y)}, \bar{\beta}_{(x,y)}, \bar{\gamma}_{(x,y)}, \bar{\delta}_{(x,y)}$ form a set of auxiliary fermions independent from $\alpha_{(x,y)},\dots,\delta_{(x,y)}$ (Fig.~\ref{FigfIso}(d)).
We define the isometric condition to be that the contraction of $\hat{A}^*$, $\hat{A}$, and the creation operators on the incoming legs give the identity tensor with auxiliary fermions attached to it. 
For example, in the upper-right quadrant (Fig.~\ref{FigfIso}(e)), the isometric condition is 
\begin{align*}
    \<\hat{A}^*\hat{A}\delta_{uu'}\gamma^{\dagger U}\bar{\gamma}^{\dagger U'}\delta_{rr'}\beta^{\dagger R}\bar{\beta}^{\dagger R'}\>
    = \delta_{dd'}\bar{\delta}^{D'}\delta^{ D}\delta_{ll'}\bar{\alpha}^{L'}\alpha^{L},
    \label{Eqfisometriccondition}
\end{align*}
where $\<\cdot\>$ denotes contraction over $\gamma,\bar\gamma,\beta,\bar\beta$, and $c$.
See an alternative explanation based on swap gates in SM~\cite{SM}.

Proof: the tensor network representations of $|\psi\>$ and $\<\psi'|$ are
\begin{align}
    |\psi'\> = \< \hat{\Lambda}'_{(x_0,y_0)}\prod_{(x,y)\neq(x_0,y_0)} \hat{A}_{(x,y)}\prod_{(x,y)}\hat{V}_{(x,y)}\>_\text{aux}|vac\rangle,\\
    \<\psi| = \<vac|\< \hat{\Lambda}^*_{(x_0,y_0)}\prod_{(x,y)\neq(x_0,y_0)} \hat{A}^*_{(x,y)}\prod_{(x,y)}\hat{V}^*_{(x,y)}\>_\text{aux},\label{Eqfermionbra}\\
    \hat{V}^*_{(x,y)} \equiv(1+\bar{\alpha}_{(x+1,y)}^{\dagger}\bar{\beta}_{(x,y)}^{\dagger})(1+\bar{\gamma}_{(x,y)}^{\dagger}\bar{\delta}_{(x,y+1)}^{\dagger})
\end{align}
To get Eq.~\ref{Eqfermionbra} we took the conjugate of Eq.~\ref{EqfPEPS} and made a particle-hole transformation of the auxiliary fermions. 
Consider contracting the fermionic tensors and the bond operators in the expression of $\<\psi|\psi'\>$ from the lower-left corner toward the orthogonality center. 
The tensor at the lower-left corner has trivial left and down legs, $l=L=0, d=D=0$; the definition of the isometric condition guarantees that contracting the tensor with its conjugate (summing over the physical index) gives $\delta_{uu'}\bar{\gamma}_1^{U'}\gamma_1^{ U}\delta_{rr'}\bar{\beta}_1^{R'}\beta_1^{R}$. 
Then contracting $\delta_{rr'}\bar{\beta}_1^{R'}\beta_1^{R}$ with the bond operator on its right gives
$\<\delta_{rr'}\bar{\beta}_1^{R'}\beta_1^{R}(1+\beta_1^\dagger\alpha_2^{\dagger})(1+\bar{\alpha}_1^{\dagger}\bar{\beta}_2^\dagger)\>_{\beta_1,\bar{\beta}_1} = \delta_{l,l'}\alpha_2^{\dagger L}\bar{\alpha}_2^{\dagger L'}$, which contracts with the second tensor on the bottom row and its conjugate. We get the desired result by continuing this process iteratively from the 4 corners to the center.
$\square$

For the computation of correlation functions on the OC, the same argument shows that tensors outside can be trivially contracted and the computation reduces to a contraction along the OC, as in the bosonic case.

\section{Replacing multiple entangled fermion pairs by a single pair}
\label{appendix:combinelegs}

\begin{figure}[htb]
\begin{center}
\includegraphics[width=0.30\textwidth]{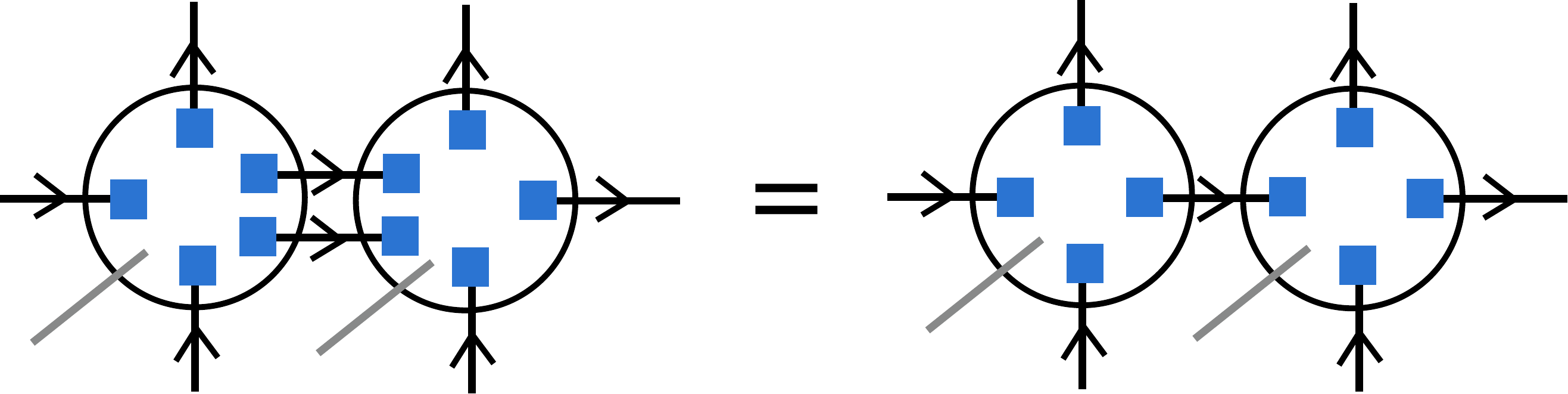}
\caption{LHS: Site 1 (left) and Site 2 (right) connected by 2 fermionic bonds with indices $i_1$ (top) and $i_2$ (bottom). RHS: Site 1 and Site 2 share a single bond with the index $i$}
\label{Figflegs}
\end{center}
\end{figure}

Consider the following ansatz where Site 1 and Site 2 share $n$ bonds $i_1\dots i_n$, with parity label $I_1\dots I_n$ (Fig.~\ref{Figflegs}).

\begin{align}
    |\psi\> &= \< \hat{A}[1]\hat{A}[2]\cdots (1+\beta_{1,1}^{\dagger}\alpha_{2,1}^{\dagger})\cdots (1+\beta_{1,n}^{\dagger}\alpha_{2,n}^{\dagger})\>_\text{aux}|vac\>\\
    \hat{A}[1] &= A[1]^{k_1}_{l_1i_1\dots i_n d_1u_1} c_1^{\dagger k_1}\delta_1^{D_1}\beta_{1,n}^{I_n}\cdots\beta_{1,1}^{I_1}\gamma_{1}^{U_1}\alpha_1^{L_1},\\
    \hat{A}[2] &= A[2]^k_{i_1\dots i_n r_2d_2u_2} c_2^{\dagger k_2}\delta_2^{D_2}\beta_{2}^{R_2}\gamma_{2}^{U_2}\alpha_{2,1}^{I_1}\cdots\alpha_{2,n}^{I_n},\\
\end{align}

We claim that replacing the n bonds with a single bond labeled by $i$ (the collection of indices $i_1\dots i_n$) and a parity label $I=I_1 + \dots + I_n$ gives the same physical state:

\begin{align}
   |\psi\> &= \< \hat{A}'[1]\hat{A}'[2]\cdots (1+\beta_{1}^{\dagger}\alpha_{2}^{\dagger})\>_\text{aux}|vac\> 
\end{align}
The new tensors are simply defined as $\hat{A}'[1] = A[1]^{k_1}_{l_1id_1u_1}c_1^{\dagger k_1}\delta_1^{D_1}\beta_{1}^{I}\gamma_{1}^{U_1}\alpha_1^{L_1} = A[1]^{k_1}_{l_1i_1\dots i_n d_1u_1}c_1^{\dagger k_1}\delta_1^{D_1}\beta_{1}^{I}\gamma_{1}^{U_1}\alpha_1^{L_1}$, $\hat{A}'[2] = A[2]^{k_2}_{ir_2d_2u_2}c_1^{\dagger k_2}\delta_2^{D_2}\beta_{2}^{R_2}\gamma_{2}^{U_2}\alpha_2^{I}$. This equivalence follows from the fact that $\beta_{1}^{I}$ ($\alpha_{2}^{I}$) has the same fermion parity as $\beta_{1,n}^{I_n}\cdots\beta_{1,1}^{I_1}$ ($\alpha_{1,1}^{I_1}\cdots\alpha_{1,n}^{I_n}$) and they give the same result after contracting with the corresponding creation operators
\begin{align}
    \< \alpha_{2}^{I}\beta_{1}^{I}(1+\beta_{1}^\dagger\alpha_2^{\dagger})\>_{\beta_1,\alpha_2} &= 1,\ \forall I\\
    \< \alpha_{2,1}^{I_1}\cdots\alpha_{2,n}^{I_n}\beta_{1,n}^{I_n}\cdots \beta_{1,1}^{I_1}(1+\beta_{1,1}^\dagger\alpha_{2,1}^{\dagger})\cdots(1+\beta_{1,n}^\dagger\alpha_{2,n}^{\dagger})\>_{\beta_{1,1},\alpha_{2,1},\cdots\beta_{1,n},\alpha_{2,n}} &= 1,\ \forall I_1,\dots ,I_n
\end{align}
Note that this result relies on the counterclockwise ordering convention of fermion operators.

\section{Equivalence of the swap gate ansatz and the virtual fermion ansatz}
\label{appendix:swapgate}

We provide a self-contained introduction to the swap-gate ansatz and show the equivalence to the virtual-fermion ansatz by explicitly contracting the virtual fermions.

\begin{figure}[htb]
\begin{center}
\includegraphics[width=0.50\textwidth]{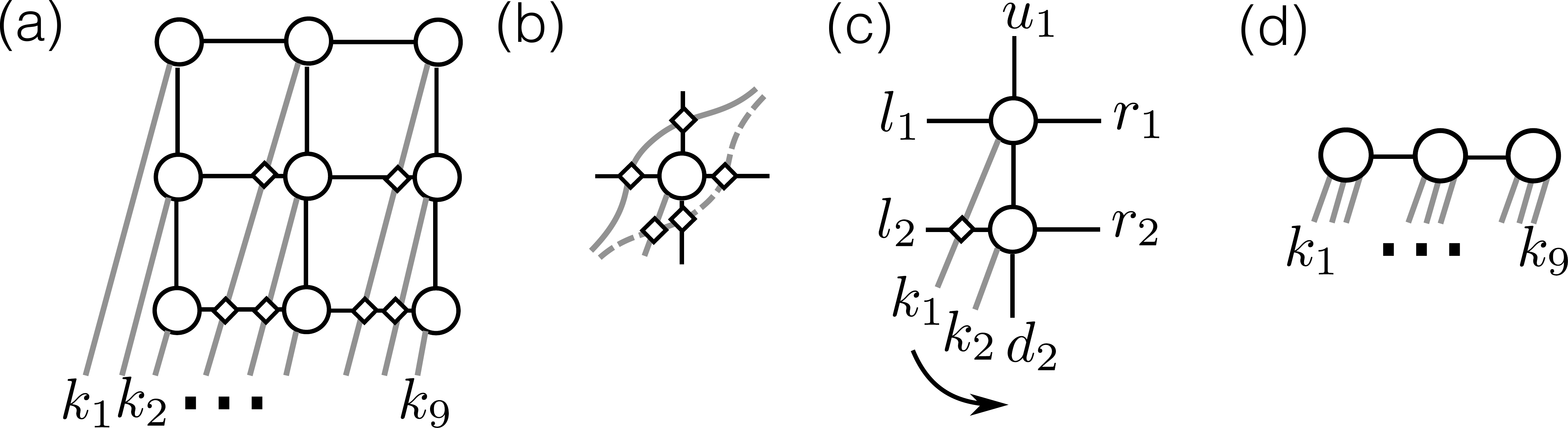}
\caption{(a) Representation of the wavefunction in the swap gate convention. The wavefunction is given by the contraction of all tensors and swap gates. (b) Deforming the physical legs give does not affect the wavefunction. (c) The contraction of two tensors. (d) The result of contracting the three tensors in each column.}
\label{Figswap}
\end{center}
\end{figure}

The swap-gate ansatz is defined by 2 rules. 
First, each tensor is $\mathbb{Z}_2$-graded and has zero total parity: each virtual index has a parity label $0$ or $1$, and the tensor element is zero unless the corresponding labels on all legs add to an even number. 
Second, whenever two lines with parity $I$ and $J$ cross each other, a swap gate $(-1)^{IJ}$ is added.
The wavefunction $\psi^{k_1\dots,k_N}$ in the basis of $c_1^{\dagger k_1}c_2^{\dagger k_2}\cdots c_N^{\dagger k_N}|vac\>$ is defined as the contraction of all tensors and swap gates as shown in Fig.~\ref{Figswap}(a).
In Fig.~\ref{Figswap}(a), the order of the endpoints of physical lines from left to right corresponds to the fermion order of the Fock-space basis. 
If we were to swap the endpoints of two physical fermion lines, for example, $k_2$ and $k_3$, we would add a swap-gate $(-1)^{k_2k_3}$. 
It is easy to see the resulting wavefunction is the wavefunction of the same state written in a different basis $c_{1}^{\dagger k_1}c_{3}^{\dagger k_3}c_{2}^{\dagger k_2}\cdots c_{N}^{\dagger k_N}|vac\>$.
On the other hand, Deforming the physical lines without changing the order of endpoints does not change the wavefunction. 
For example, in Fig.~\ref{Figswap}(b), the physical line $k_1$ may pass through the tensor $A^k_{lrdu}$ (parity label $k, L, R, D, U$) in two ways (the solid line and the dashed line). 
For the solid line, there are 2 swap gates giving $(-1)^{k_1(L+U)}$; for the dashed line, there are 3 swap gates giving $(-1)^{k_1(k+D+R)}$. Since the tensor has even parity, the tensor elements are zero unless $L+U = k+D+R\ (mod\ 2)$. 
Thus, the two choices of swap gates produce the same wavefunction.

Now we show the virtual-fermion ansatz we discussed in the main text produces the same wavefunction by explicitly contracting all virtual fermions. We first contract all tensors in each column, and then contract the resulting tensors of different columns.
As shown in Fig.~\ref{Figswap}(c), we can replace the top tensor ($A[1]$) and the bottom tensor ($A[2]$) with a single tensor ($\Theta$, fermion operators ordered counterclockwise starting from physical legs). Using the virtual-fermion ansatz, we have

\begin{align}
    \hat{\Theta} = \sum_{r1=l2}\< 
    A[2]^{k_2}_{l_2r_2d_2i}
    c_2^{\dagger k_2}\delta_2^{D_2}\beta_{2}^{R_2}\gamma_{2}^{I}\alpha_2^{L_2}
    A[1]^{k_1}_{l_1r_1iu_1}  
    c_1^{\dagger k_1}\underline{\delta_1^{I}\beta_{1}^{R_1}\gamma_{1}^{U_1}\alpha_1^{L_1}}(1+\delta_1^{\dagger}\gamma^{\dagger}_2)\>_{\delta_1,\gamma_2}
    \label{Eqappendixupdown}\\
    \hat{\Theta} \equiv\Theta^{k_1k_2}_{l_1l_2r_1r_2d_2u_1}c_1^{\dagger k_1}c_2^{\dagger k_2}\delta_2^{D_2}\beta_2^{R_2}\beta_1^{R_1}\gamma_1^{U_1}\alpha_1^{L_1}\alpha_2^{L_2}
    \label{Eqappendixtheta}
\end{align}
We bring the RHS of Eq.~\ref{Eqappendixupdown} to the form of \ref{Eqappendixtheta} in 3 steps: bringing $c_1^{\dagger k_1}$ to the left, moving $\alpha_2^{L_2}$ to the right, and contracting $\gamma_2^I\delta_1^I$ with the creation operator. 
In the second step, $\alpha_2^{L_2}$ is swapped with the underlined operators, with total parity $k_1$, hence the sign $(-1)^{k_1L_2}$ in Fig.~\ref{Figswap}.
We contract tensors in a column from the top to the bottom. Each physical leg needs to pass through all horizontal legs on the left of the column below it, hence the minus signs indicated by the swap gates in Fig.~\ref{Figswap}(a). After the vertical contractions, we get a tensor for each column (Fig.~\ref{Figswap}(d)). It is easy to see that contracting them does not give any extra sign.

The argument above shows the equivalence of the virtual-fermion ansatz and the swap-gate ansatz in the case that the physical legs are extended outside as in Fig.~\ref{Figswap}(a). Using the consistency relations discussed earlier in this section, this equivalence stands no matter how we extend the physical legs outside.

\section{Alternative interpretation of the isometric condition in the swap-gate convention}

In the swap-gate convention, it's important to fix an order of the legs for tensors representing the state $|\psi\>$. 
We choose to the order in Fig.~\ref{Figswapiso}(a). 
Counting counterclockwise from the physical leg, the order is `physical, down, right, up, left', the same as the order of fermion operators in the virtual-fermion ansatz.
The order of legs for tensors representing the conjugate $\<\psi|$ must be the opposite as the order for tensors representing $|\psi\>$ (Fig.~\ref{Figswapiso}(b)). This is in analogy of the reverse of orders when taking the conjugate of products of operators.

\begin{figure}[htb]
\begin{center}
\includegraphics[width=0.40\textwidth]{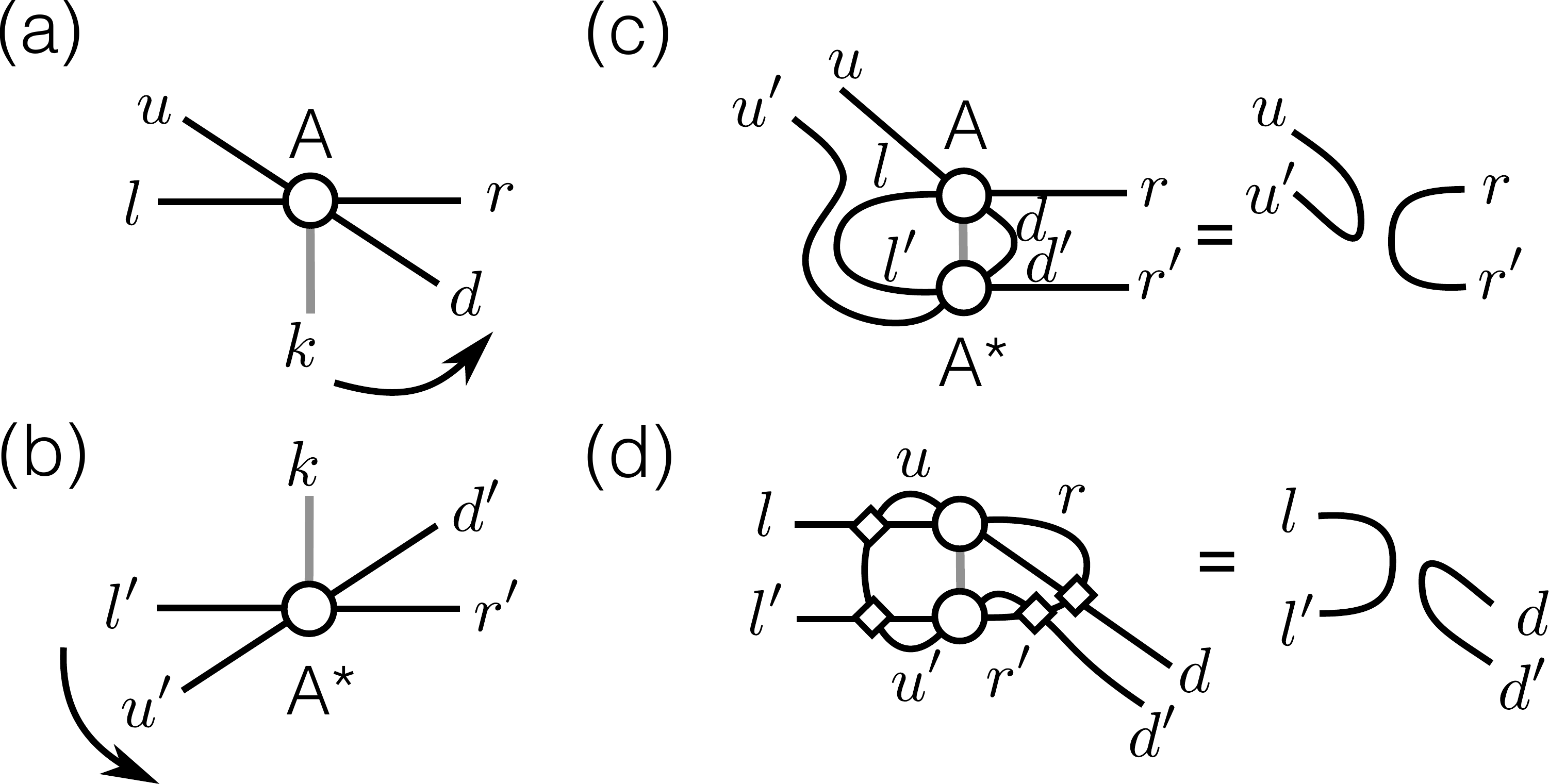}
\caption{(a) The legs of tensors representing $|\psi\>$ are ordered counterclockwise according to `physical, down, right, up, left'. (b) The legs of tensors representing $\<\psi|$ have the opposite order. (c) The isometric condition for fTNS in the lower-left quadrant. (d) The isometric condition for fTNS in the upper-right quadrant.}
\label{FigSupswapiso}
\end{center}
\end{figure}

With this convention, we can give a pictorial definition of the isometric condition for fermionic tensor networks.
The isometric condition is shown in Fig.~\ref{FigSupswapiso}(c) for tensors in the lower-left quadrant and in Fig.~\ref{FigSupswapiso}(d) for tensors in the upper-right quadrant.
In the lower-left quadrant, if contracting the left, physical, and down legs gives delta functions in the right and up legs, the network can be contracted trivially from the lower-left corner to the central cross. Since there is no cross of the legs, the isometric condition is simply $\sum_{k,l,d}A^k_{lrdu}A^{*k}_{lr'du'} = \delta_{uu'}\delta_{rr'}$, namely $A$ is an isometry.

On the other hand, in the upper-right quadrant, we need to contract tensors from the upper-right to the lower-left. The contraction of the up and right legs leads to 4 swap gates, hence the isometric condition $\sum_{k,r,u}(-1)^{u(L+L')+r(D+D')}A^{k}_{lrdu}A^{*k}_{l'rd'u} = \delta_{ll'}\delta_{dd'}$. This condition is equivalent to that $(-1)^{UL+RD}A$ is an isometry. We can equivalently write this condition as $(-1)^{UD+RD}A$ is an isometry, by changing $L$ into $D+U+k+R$ (total parity even) and using the fact that $(-1)^{U(k+U+R)}$ is a unitary transformation on the incoming legs (diagonal matrix with $\pm 1$ diagonal elements). Finally, we can simplify the isometric condition to that $(-1)^{kD} A$ is an isometry by changing $U+R$ to $k+D+L$ and using the fact that $(-1)^{D(D+L)}$ is a unitary on the outgoing legs.

A similar argument shows that the isometric condition in the lower-right and upper-left quadrants is also that $A$ is an isometry from outgoing legs to incoming legs.

\section{Finite-depth-local-unitary circuits preserve fisoTNS}

We prove that if a fermionic state has a finite-bond-dimension fisoTNS representation in the thermodynamic limit, it can still be represented to any accuracy by a finite-bond-dimension fisoTNS after the action of a finite-depth-local-unitary (FDLU) circuit.

\begin{figure}[htb]
\begin{center}
\includegraphics[width=0.7\textwidth]{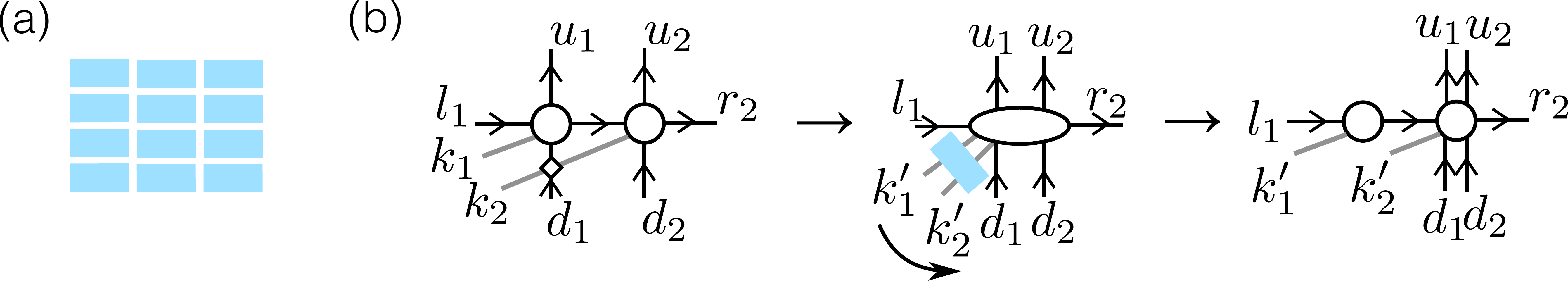}
\caption{(a) One layer of local 2-site unitary gates. (b) The process of getting the fisoTNS representation of the final state in the lower-left quadrant}
\label{Figcombinesplit}
\end{center}
\end{figure}
Since any FDLU can be deformed into finite-depth 2-site unitary gates to any desired accuracy, it suffices to derive an fisoTNS representation after a single layer of 2-site unitaries (Fig.~\ref{Figcombinesplit}).
We demonstrate it in the lower-left quadrant. 

To do so, we combine the 2 tensors on the neighboring sites and act it by the corresponding unitary gate.

\begin{align}
    \hat{\Theta} = \<
    A[1]c_1^{\dagger k_1}\delta_1^{D_1}\beta_{1}^{I}\gamma_{1}^{U_1}\alpha_1^{L_1}
    A[2]\underline{c_2^{\dagger k_2}\delta_2^{D_2}\beta_{2}^{R_2}\gamma_{2}^{U_2}\alpha_2^{I}}(1+\beta^{\dagger}_1\alpha^{\dagger}_2)\>_{\beta_1,\alpha_2}
    \\
    \hat{\Theta} \equiv \Theta^{k_1k_2}_{l_1r_2d_1d_2u_1u_2}c_1^{\dagger k_1}c_2^{\dagger k_2}\delta_1^{D_1}\delta_2^{D_2}\beta_2^{R_2}\gamma_2^{U_2}\gamma_1^{U_1}\alpha_1^{L_1}
\end{align}
To derive an expression for tensor $\Theta$, we place the underlined operators between $\delta_1$ and $\beta_1$, contract $\alpha_2^{I}\beta_{1}^I$ with the creation operator, and then swap $\delta_1^{D_1}$ and $c_2^{k_2}$. Since the underlined operators have even total parity, moving it does not give a minus sign; the only step that may produce a minus sign is the swap. 
Thus we have $\Theta = A[1]A[2](-1)^{k_2D_1}$. Note that if $A[1]$ and $A[2]$ are isometries with respect to the arrows in Fig.~\ref{Figcombinesplit}, The contraction $A[1]A[2]$ is the composition of two isometries, which is still an isometry. 
$\Theta$ is also an isometry since the extra factor $(-1)^{k_2D_1}$ is a unitary transformation on the incoming legs.
For the same reason, the physical gate acting on the leg $k_1$ and $k_2$, $\Theta' = U\Theta$, also preserves the isometric structure.

Then we split the tensor into two isometric tensors as illustrated in Fig.~\ref{Figcombinesplit}(b). 
The internal bond is just the combination of $k'_1$ and $l_1$. 
The left tensor is the identity matrix from the outgoing leg to the incoming legs, with trivial up and down legs. 
The right tensor is $\Theta'$.

With minor changes regarding the corresponding definition of isometric form, the same procedure applies to other quadrants and the central cross.

\section{Moses Move for symmetric TNS}
The MM algorithm for isoTNS without symmetry is described in \cite{zaletel2020isometric}. 
We repeat it here. 
The MM algorithm performs the following decomposition of the orthogonality column (OC) into a left-canonical isometric column and a new OC:
\begin{equation}
    \label{eq:MM}
    \newcommand{\LL}{3}      
\renewcommand{\d}{1.0}   
\renewcommand{\r}{0.1}   
\renewcommand{\a}{0.8}   
\newcommand{\dH}{0.5}   
\newcommand{\aH}{0.3}   
\begin{tikzpicture}[baseline = (X.base),every node/.style={scale=1.0},scale=1.0]
\newcommand{\x}{0}
\newcommand{\y}{0}
\draw (\LL*\d/2,\LL*\d/2) node (X) {};
\foreach \j in {0,...,\LL}
{
  \renewcommand{\x}{0}
  \renewcommand{\y}{\j*\d}
  \pgfmathsetmacro{\color}{ifthenelse(\j==0, "red", "blue")}
  \draw [fill=\color] (\x,\y) circle (\r);
  \draw [midarrow={latex reversed}](\x+\r,\y) -- (\x+\r+\aH,\y); 
  \draw [midarrow={latex}](\x-\r-\aH,\y) -- (\x-\r,\y); 
  \ifthenelse{\j<\LL}{
  \draw [midarrow={latex reversed}](\x,\y+\r) -- (\x,\y+\r+\a);
  }{}
}
\draw (0,-0.5) node {(i)};
\end{tikzpicture}
\hspace{-12mm}
\approx
\hspace{-3mm}
\begin{tikzpicture}[baseline = (X.base),every node/.style={scale=1.0},scale=1.0]
\newcommand{\x}{0}
\newcommand{\y}{0}
\draw (\LL*\d/2,\LL*\d/2) node (X) {};
\foreach \j in {1,...,\LL}
{
  \renewcommand{\x}{0}
  \renewcommand{\y}{\j*\d}
  \pgfmathsetmacro{\color}{ifthenelse(\j==0, "red", "blue")}
  \draw [fill=\color] (\x,\y) circle (\r);
  \draw [midarrow={latex reversed}](\x+\r,\y) -- (\x+\r+\aH,\y); 
  \draw [midarrow={latex}](\x-\r-\aH,\y) -- (\x-\r,\y); 
  \ifthenelse{\j<\LL}{
  \draw [midarrow={latex reversed}](\x,\y+\r) -- (\x,\y+\r+\a);
  }{}<++>
}
\renewcommand{\x}{-\dH/2}
\renewcommand{\y}{0}
\draw [midarrow={latex reversed}](\x-\r,\y) -- (\x-\r-\aH,\y); 
\draw [fill=black] (\x,\y) circle (\r);
\draw [midarrow={latex}](\x+\r,\y) -- (\x+\r+\aH,\y); 
\renewcommand{\x}{\dH/2}
\draw [fill=blue] (\x,\y) circle (\r);
\draw [midarrow={latex reversed}](\x+\r,\y) -- (\x+\r+\aH,\y); 

\renewcommand{\x}{0}
\renewcommand{\y}{\d*0.5}
\draw [fill=red] (\x,\y) circle (\r);
\draw [midarrow={latex reversed}](\x,\y+\r) -- (\x,\y-\r+\d*0.5); 
\renewcommand{\x}{-\dH/2}
\renewcommand{\y}{0}
\draw [midarrow={latex}](\x+\r*0.5,\y+\r*0.866) -- (0-\r*0.5,\y+\d*0.5-\r*0.866); 
\renewcommand{\x}{\dH/2}
\draw [midarrow={latex}](\x-\r*0.5,\y+\r*0.866) -- (0+\r*0.5,\y+\d*0.5-\r*0.866); 
\draw (0,-0.5) node {(ii)};
\end{tikzpicture}
\hspace{-11mm}
=
\hspace{-1mm}
\begin{tikzpicture}[baseline = (X.base),every node/.style={scale=1.0},scale=1.0]
\newcommand{\x}{0}
\newcommand{\y}{0}
\draw (\LL*\d/2,\LL*\d/2) node (X) {};
\foreach \j in {1,...,\LL}
{
  \renewcommand{\x}{0}
  \renewcommand{\y}{\j*\d}
  \pgfmathsetmacro{\color}{ifthenelse(\j==1, "red", "blue")}
  \draw [fill=\color] (\x,\y) circle (\r);
  \draw [midarrow={latex reversed}](\x+\r,\y) -- (\x+\r+\aH,\y); 
  \draw [midarrow={latex}](\x-\r-\aH,\y) -- (\x-\r,\y); 
  \ifthenelse{\j<\LL}{
  \draw [midarrow={latex reversed}](\x,\y+\r) -- (\x,\y+\r+\a);
  }{}<++>
}
\renewcommand{\x}{-\dH/2}
\renewcommand{\y}{0}
\draw [midarrow={latex reversed}](\x-\r,\y) -- (\x-\r-\aH,\y); 
\draw [fill=black] (\x,\y) circle (\r);
\draw [midarrow={latex}](\x+\r,\y) -- (\x+\r+\aH,\y); 
\renewcommand{\x}{\dH/2}
\draw [fill=blue] (\x,\y) circle (\r);
\draw [midarrow={latex reversed}](\x+\r,\y) -- (\x+\r+\aH,\y); 

\renewcommand{\x}{0}
\renewcommand{\y}{\d*0.5}
\renewcommand{\x}{-\dH/2}
\renewcommand{\y}{0}
\draw [midarrow={latex}](\x+\r*0.24,\y+\r*0.866) -- (0-\r*0.24,\y+\d-\r*0.866); 
\renewcommand{\x}{\dH/2}
\draw [midarrow={latex}](\x-\r*0.24,\y+\r*0.866) -- (0+\r*0.24,\y+\d-\r*0.866); 
  \draw (0,-0.5) node {(iii)};
\end{tikzpicture}
\hspace{-11mm}
\approx
\hspace{-1mm}
\begin{tikzpicture}[baseline = (X.base),every node/.style={scale=1.0},scale=1.0]
\newcommand{\x}{0}
\newcommand{\y}{0}
\draw (\LL*\d/2,\LL*\d/2) node (X) {};
\foreach \j in {2,...,\LL}
{
  \renewcommand{\x}{0}
  \renewcommand{\y}{\j*\d}
  \pgfmathsetmacro{\color}{ifthenelse(\j==2, "red", "blue")}
  \draw [fill=\color] (\x,\y) circle (\r);
  \draw [midarrow={latex reversed}](\x+\r,\y) -- (\x+\r+\aH,\y); 
  \draw [midarrow={latex}](\x-\r-\aH,\y) -- (\x-\r,\y); 
  \ifthenelse{\j<\LL}{
  \draw [midarrow={latex reversed}](\x,\y+\r) -- (\x,\y+\r+\a);
  }{}
}
\foreach \j in {0,...,1}
{
  \renewcommand{\x}{-\dH/2}
  \renewcommand{\y}{\j*\d}
  \draw [midarrow={latex reversed}](\x-\r,\y) -- (\x-\r-\aH,\y); 
  \draw [fill=black] (\x,\y) circle (\r);
  \draw [midarrow={latex}](\x+\r,\y) -- (\x+\r+\aH,\y); 
  \renewcommand{\x}{\dH/2}
  \draw [fill=blue] (\x,\y) circle (\r);
  \draw [midarrow={latex reversed}](\x+\r,\y) -- (\x+\r+\aH,\y); 
  \ifthenelse{\j<1}{
  \foreach \i in {0,...,1}
  {
  \renewcommand{\x}{-\dH/2+\i*\dH}
  \draw [midarrow={latex}](\x,\y+\r) -- (\x,\y+\r+\a);
  }
  }{}
}
\renewcommand{\x}{-\dH/2}
\renewcommand{\y}{\d}
\draw [midarrow={latex}](\x+\r*0.24,\y+\r*0.866) -- (0-\r*0.24,\y+\d-\r*0.866); 
\renewcommand{\x}{\dH/2}
\draw [midarrow={latex}](\x-\r*0.24,\y+\r*0.866) -- (0+\r*0.24,\y+\d-\r*0.866); 
  \draw (0,-0.5) node {(iv)};
\end{tikzpicture}
\hspace{-13mm}
\approx
...
\approx
\hspace{-3mm}
\begin{tikzpicture}[baseline = (X.base),every node/.style={scale=1.0},scale=1.0]
\draw (\LL*\d/2,\LL*\d/2) node (X) {};
\newcommand{\x}{0}
\newcommand{\y}{0}
\foreach \i in {0,...,1}
{
  \foreach \j in {0,...,\LL}
  {
    \renewcommand{\x}{\i*\dH}
    \renewcommand{\y}{\j*\d}
    \pgfmathsetmacro{\Hdir}{ifthenelse(\i == 0, "latex", "latex reversed")}
    \pgfmathsetmacro{\Vdir}{ifthenelse(\j <\LL,"latex", "latex reversed")}
    \pgfmathsetmacro{\color}{ifthenelse(\i==0, ifthenelse(\j==\LL, "blue", "black"), ifthenelse(\j<\LL, "blue", "red"))}
    \draw [fill=\color] (\x,\y) circle (\r);
    \draw [midarrow={\Hdir}](\x+\r,\y) -- (\x+\r+\aH,\y); 
    \ifthenelse{\j<\LL}{
    \draw [midarrow={\Vdir}](\x,\y+\r) -- (\x,\y+\r+\a);
    }{}
  }
}
\foreach \j in {0,...,\LL}
{
  \renewcommand{\x}{0*\d}
  \renewcommand{\y}{\j*\d}
  \draw [midarrow={latex reversed}](\x-\r,\y) -- (\x-\r-\aH,\y); 
}
  \draw (\dH/2,-0.5) node {(v)};
\end{tikzpicture}
    \hspace{-1cm}
\end{equation}
where the physical legs of the isoTNS are grouped either with the left or the right virtual index.
The MM algorithm effectively ``unzips" the original OC by iteratively splitting the (red) 5-leg tensor at the orthogonality center into 3 smaller tensors.

Because of the isometric form, 
each splitting step in the MM
can be reduced to the following local tripartite decomposition:   
\begin{equation}
  \label{eq:tripartite}
\newcommand{\x}{0}
\newcommand{\y}{0}
\renewcommand{\d}{1.0}   
\renewcommand{\r}{0.1}   
\renewcommand{\a}{0.8}   
\begin{tikzpicture}[baseline = (X.base),every node/.style={scale=1.0},scale=1.0]
\draw (0,0) node (X) {};
  \draw [fill=red] (\x,\y) circle (\r);
  \draw [midarrow={latex reversed}](\x+\r,\y) -- (\x+\r+\a,\y); 
  \draw [midarrow={latex}](\x-\r-\a,\y) -- (\x-\r,\y); 
  \draw [midarrow={latex reversed}](\x,\y+\r) -- (\x,\y+\r+\a);
\end{tikzpicture}
\approx
\begin{tikzpicture}[baseline = (X.base),every node/.style={scale=1.0},scale=1.0]
  \newcommand{\dH}{0.8}
  \renewcommand{\a}{0.6}
  \draw (\x-\d*1.3,\y) node{$f$};
  \draw (\x+\d*1.3,\y) node{$e$};
  \draw (\x+0.2,\y+\d*1.2) node{$d$};
  \draw (0,-0.25) node{$b$};
  \draw (-\dH/4*1.7,\dH/2*1.1) node{$c$};
  \draw (\dH/4*1.7,\dH/2*1.1) node{$a$};
  \renewcommand{\x}{-\dH/2}
  \renewcommand{\y}{0}
  \draw [fill=black] (\x,\y) circle (\r);
  \draw [midarrow={latex reversed}] (\x-\r,\y) -- (\x-\r-\a,0);
  \draw [midarrow={latex}] (\x+\r,\y) -- (\x+\r+\a,0);
  \draw [midarrow={latex}] (\x+\r*0.5,\y+\r*0.866) -- (0-\r*0.5,\dH*0.866-\r*0.866);
  \renewcommand{\x}{\dH/2}
  \renewcommand{\y}{0}
  \draw [fill=blue] (\x,\y) circle (\r);
  \draw [midarrow={latex reversed}] (\x+\r,\y) -- (\x+\r+\a,0);
  \draw [midarrow={latex}] (\x-\r*0.5,\y+\r*0.866) -- (0+\r*0.5,\dH*0.866-\r*0.866);
  \renewcommand{\x}{0}
  \renewcommand{\y}{\dH*0.866}
  \draw [fill=red] (\x,\y) circle (\r);
  \draw [midarrow={latex reversed}] (\x,\y+\r) -- (\x,\y+\r+\a);
\end{tikzpicture}. 
\end{equation}
where we view the five-leg red tensor in (iii) of Eq. \ref{eq:MM} as a tripartite tensor by grouping its lower left and lower right index respectively with its left and right index.  
Typically, one needs to place a limit on the maximum internal bond dimension of the decomposition, which generally disallows an exact equality in Eq. \ref{eq:tripartite}.  
This decomposition can be done as follows.  
\begin{enumerate}
  \item One first groups index $d$ and $e$ and then performs an SVD decomposition on the grouped matrix $\Psi_{f:de} = UsV^\dag$. 
    The index between $U$ and $sV^\dag$, arising from the SVD decomposition, is split into indices $b$ and $c$.  
    The permutational freedom in splitting the indices will be fixed by the disentangler $u$ in step 2. 
    The bond dimensions of $b$ and $c$ are truncated when necessary. 
  \item Note that the SVD in step 1 has a unitary freedom, $u$, acting on the indices $b$ and $c$: $UsV^\dag = Uuu^\dag s V^\dag$.
    To make use of this freedom, the tensor $u^\dag s V^\dag$ is reshaped into a matrix $\Theta$ with its $b$ and $e$ indices grouped as the column index and the $c$ and $d$ grouped as the row index.
    $u$ is chosen so that the Renyi-$\alpha$ entropy of the matrix $\Theta$ is minimized, for a certain Renyi index $\alpha$. 
    We choose $\alpha=1/2.$
    The disentangler, $u$, is optimized in order to minimize the error of the SVD truncation in step 3. 
The black tensor is then equal to the reshaped $Uu$. 
\item Perform an SVD on $\Theta$: $\Theta=U's'V'^\dag$.   
The blue tensor is then equal to the reshaped $V'^\dag$ after truncation, and the red tensor is equal to the truncated $U's'$.  
\end{enumerate}

In the presence of \ZZ\, symmetry, all operations described above respect the charge structure naturally, except that in step 1, when the index between $U$ and $sV^\dag$ is split into indices $b$ and $c$, a charge structure on $b$ and $c$ has to be prescribed.
Let $n_0$ and $n_1$ respectively be the size of the charge-0 and charge-1 block on the bond between $U$ and $sV^\dag$ before the splitting.
Then we have to prescribe the size of the charge-0 and charge-1 block on both bond $b$ and $c$. 
We denote them as $n_0^b, n_1^b, n_0^c,$ and $ n_1^c$. They have to satisfy the constraint
\begin{align}
    n_0^b n_0^c + n_1^b n_1^c \le n_0 \\
    n_0^b n_1^c + n_1^b n_0^c \le n_1
\end{align}
The charge block of size $n_0 n_1 - (n_0^b+n_1^b)(n_0^c+n_1^c)$ will need to be truncated. 
We choose the $n$s such that the size of the truncated block is the smallest. 
When multiple sets of $n$s tie, we choose the set for which the standard deviation of the list $[n_0^b, n_1^b, n_0^c, n_1^c]$ is the least, i.e. we choose the evenest charge block sizes. 
In the code, $[n_0^b, n_1^b, n_0^c, n_1^c]$ is chosen via explicit enumeration and comparison. 
The computational cost for this is negligible. 
\end{document}